\documentclass[lettersize,journal]{IEEEtran}
\usepackage{multirow}
\usepackage{amsfonts,amsmath,amssymb,mathtools}
\usepackage{graphicx}
\usepackage{epstopdf}
\usepackage{xurl}
\usepackage{cite}
\usepackage[caption=false,font=footnotesize,subrefformat=parens,labelformat=parens]{subfig}
\usepackage{comment}
\usepackage[ruled,vlined]{algorithm2e}
\usepackage{algpseudocode}
\usepackage{tikz}
\usetikzlibrary{arrows,intersections,decorations,plotmarks,math,patterns}
\usepackage{pgfplots}
\pgfplotsset{compat=1.17}
\usepackage{color}
\usepackage[colorinlistoftodos]{todonotes}

\definecolor{COL11}{RGB}{215,25,28}
\definecolor{COL12}{RGB}{253,174,97}
\definecolor{COL13}{RGB}{146,197,222}
\def\BibTeX{{\rm B\kern-.05em{\sc i\kern-.025em b}\kern-.08em
    T\kern-.1667em\lower.7ex\hbox{E}\kern-.125emX}}
\usepackage{balance}
\usepackage{booktabs} 
\usepackage{hyperref}
\newcommand\scalemath[2]{\scalebox{#1}{\mbox{\ensuremath{\displaystyle #2}}}}

\begin{document}

\title{Bi-Residual Neural Network based Synchronous Motor Electrical Faults Diagnosis: Intra-link Layer Design for High-frequency Features}
% 这么名字有点土
\author{
Qianchao Wang$^{1}$ \textit{Member, IEEE}, Leena Heistrene$^{2}$ \textit{Member, IEEE}, Yoash Levron$^{3}$ \textit{Senior Member, IEEE}, Yuxuan Ding$^{1\dag}$, Yaping Du $^{1}$,

\thanks{Yuxuan Ding is the corresponding author }% <-this % stops a space
\thanks{$^{1}$Qianchao Wang, Yuxuan Ding, Yaping Du, are with the Department of Building Environment and Energy Engineering, Hong Kong Polytechnic University, Hung Hom, Hong Kong. (qianchao.wang@polyu.edu.hk; yx.ding@connect.polyu.hk; Ya-ping.du@polyu.edu.hk;)}
\thanks{$^{2}$Leena Heistrene is Department of Electrical Engineering, School of Energy Technology, Pandit Deendayal Energy University, Gandhinagar, India. (leena.santosh@sot.pdpu.ac.in)}
\thanks{$^{3}$Yoash Levron is with the Andrew and Erna Viterbi Faculty of Electrical \& Computer Engineering, Technion—Israel Institute of Technology, Haifa, 3200003, Israel (yoashl@ee.technion.ac.il)}
}
\maketitle

\begin{abstract}
% In synchronous motor electrical fault diagnosis, the critical information is normally confined to the high-frequency components. 
% Synchronous motors are essential in industrial applications due to their stability and efficiency across various operating conditions. 
% Effective fault diagnosis of synchronous motors is crucial to ensure reliable operation, prevent costly downtime, and enhance safety. However, in practical resource-constrained environments, efficiently extracting the potential high-frequency fault-critical information presents a significant challenge. 
In practical resource-constrained environments, efficiently extracting the potential high-frequency fault-critical information is an inherent problem. 
% Hence, the inherent problem is how to extract the information within limited computation resources automatically. 
To overcome this problem, this work suggests leveraging a bi-residual neural network named Bi-ResNet to extract the inner spatial-temporal high-frequency features using embedded spatial-temporal convolution blocks and intra-link layers. It can be considered as embedding a high-frequency extractor into networks without adding any parameters, helping shallow networks achieve the performance of deep networks. 
In our experiments, five advanced CNN-based neural networks and two baselines across a real-life dataset are utilized for synchronous motor electrical fault diagnosis to demonstrate the effectiveness of Bi-ResNet including one analytical, comparative, and ablation experiments. The corresponding experiments show: 1) The Bi-ResNet can perform better on low-resolution noisy data. 2) The proposed intra-links can help high-frequency components extraction and location from raw data. 3) There is a trade-off between intra-link number and input data complexity.

\end{abstract}


\section{Introduction}\label{Introduction}
\IEEEPARstart{I}{n} industrial applications, synchronous motors play a vital role due to their stability, efficiency, and adaptability across diverse operational conditions. However, electrical faults in these motors pose significant risks, as undetected issues can lead to operational failures, costly repairs, and safety concerns. Effective fault diagnosis in synchronous motors helps prevent unexpected downtime, reduce maintenance costs, and extend equipment lifespan, making it a critical area for industrial reliability and efficiency~\cite{bhuiyan2020survey}. This necessity drives the ongoing development of advanced diagnostic techniques that improve accuracy and robustness in detecting these faults under real-world constraints.

With the development of deep learning (DL), the neural networks have demonstrated their outstanding performance when applied to synchronous motors electrical faults diagnosis~\cite{10151958,10330029}. However, these state-of-art networks mostly rely on the data pre-processing for example discrete wavelet transform~\cite{9580592} to find the high-frequency components, improving the classification accuracy with limited computation resources. In another aspect, the performance of fault diagnosis is limited by the performance of data pre-processing, resulting in the fragility of these methods. It is necessary to establish end-to-end models to improve the robustness of the methods.

%%%%%%%%%%%%%%%%%%%%%%%%%%%%%% end-to-end review 
In this vein, in the last couple of years, some end-to-end neural networks have been suggested for synchronous motor electrical fault diagnosis which in most cases perform better than traditional classification methods. In earlier works, some basic deep learning models are roughly utilized as the classification models such as long short-term memory (LSTM)~\cite{10100645}, Bayesian networks~\cite{cai2021data}, and 1D convolution neural networks (1D-CNN)~\cite{10472098}. In these models, raw signals are directly processed searching for the necessary features, and the approximation ability of networks is fully exploited since the data pre-processing and fault diagnosis are treated as a whole.

% Due to this problem, 
% Inspired by computer science, 
Recently, some advanced network architectures are also designed for synchronous motor electrical fault diagnosis. The feasibility of Variational Autoencoder (VAE) and its variants are explored in~\cite{quseiri2024fault}. The performance of VAE is evaluated for different operating conditions consisting of different loads and speeds. Similar to ~\cite{quseiri2024fault}, stacked VAE is leveraged in custom double-sided phase space reconstruction (CDPSR) image-driven fault diagnosis for permanent magnet synchronous motor (PMSM) under few-labeled samples~\cite{10319110}. Nevertheless, the core idea of VAE is mapping the feature space to a high dimension space which is hard to understand and difficult to relate to the high-frequency component space~\cite{kingma2022autoencodingvariationalbayes}. On the contrary, some mixed deep neural networks are easy to understand, especially in feature extraction such as wavelet scattering convolution network (WSCN)~\cite{9750877}, and the transformer and attention based classifiers~\cite{10038556}.
% the combination of CNN with Gated Recurrent Unit (GRU) or LSTM~\cite{kaya2024efficient}, the transformer and attention based classifiers~\cite{10038556} and Gaussian Bayes classifier with Mahalanobis metric~\cite{yue2024relationship}. 
In these models, the potential spatial-temporal high-frequency can be captured by neurons implicitly rather than explicitly. Therefore, a question naturally comes up: \textit{How can we explicitly capture the spatial-temporal high-frequency components?}

%%%%%%%%%%%%%%%%%%%%%%%%%%%%%%%redidual理论和应用
A possible answer is residual learning or shortcut connection~\cite{he2016deep}. The essence of residual neural network (ResNet) is to force the model to learn the identity transformation~\cite{Yu_2018_CVPR}, helping models to identify the perturbation~\cite{hauser2019residualnetworkslearningperturbation} which coincides with learning the high-frequency components in raw signals. From the respective of ordinary/partial differential equations (ODE/PDE), ResNet can be recognized as the weak approximations of differential equations~\cite{Sun2018StochasticTO}. By proper training, the discrete dynamics defined by a ResNet are close to the continuous one of a Neural ODE~\cite{NEURIPS2022_ecc38927}, giving a generalization bound~\cite{NEURIPS2023_98ed250b}. This theory demonstrates that the residual block can extract the high-order information for example jacobian matrices during the training process, supplying enough necessary information for classification. Considering this situation, the residual block is a perfect candidate for synchronous motors electrical faults diagnosis since the fault information normally appears as a high-frequency signal.

The previous application research also demonstrated the efficiency of ResNet in high-frequency component capture. In image classification, ResNet with CNN tends to converge at the local optimum which is closely related to the high-frequency components of the training images~\cite{cheng2019high}, indicating that residual learning is more sensitive to high-frequency signals. In other applications, the super-resolution problem can be effectively solved by using residual learning-based CNN. The good performance is attributed not only to the depth of the network but also to its residual structure~\cite{10485711}. When comes to industrial applications, residual CNN is widely in magnetic resonance imaging~\cite{fang2024hfgn}, and synchronous motor fault diagnosis~\cite{10203017,kaya2024efficient} as well,
%bearing fault diagnosis~\cite{9851479} as well, 
regardless of model size. 
Despite the high classification accuracy, these models still rely heavily on the number of parameters, limiting the identification of high-frequency fault signals. Therefore two main question naturally come up: \textit{How to leverage residual learning in a shallow neural network to capture spatial-temporal high-frequency components for classification, especially when the computation resources are limited?} And, \textit{if the dataset is lack of sufficient high-frequency components (low resolution), can the residual learning methods work?}
%%提出的方法解决的是有限资源下的显性高频提取问题

In this light, the primary objective of this work is fully utilizing the ability of high-frequency components learning of residual block to establish a novel and elegant shallow neural network model within limited computational resources especially when the dataset is low-resolution. We proposed a novel bi-residual neural network by designing an embedded spatial-temporal convolution block and hierarchically linking the inner neurons at the same layer in residual blocks, termed Bi-ResNet. Different from \cite{chen2021amplitude}, the spatial-temporal high-frequency components can be learned by both multi-level internal and external residual blocks with multi-receptive fields, using the union input signals. Notably, this new architecture holds the potential for adaptation to other domains within industrial diagnostic applications or regression modeling. To demonstrate the effectiveness of Bi-ResNet, Five advanced CNN-based models and two baselines are compared using high-resolution and low-resolution motor fault datasets with multi-level noise. The corresponding experiments are utilized to explore the scalability of intra-links, the importance of features, and the influence of the number of intra-links. In summary, the key contributions of this paper encompass:
\begin{enumerate}
	\item A novel high-frequency learning neural network named Bi-ResNet is designed for synchronous motor electrical fault diagnosis by using internal and external residual blocks. It helps shallow models capture the spatial-temporal high-frequency components for classification when the computation resources are limited.
        \item The embedded spatial-temporal convolution block consisting of multiple convolution kernels and the intra-linked layers consisting of inner residual learning are introduced into the neural networks to help shallow networks achieve the performance of deep networks.
        \item We conduct an experiment to generate numerous synchronous motor electrical failure data. Five advanced CNN-based deep learning models and two baselines are tested across high-resolution and low-resolution motor electrical fault datasets with multiple levels of noise to validate the effectiveness of Bi-ResNet.
        \item The experiments demonstrate the importance of intra-links in high-frequency component extraction and location. The trade-off between intra-link number $n$ and input data complexity is explored in ablation experiments as well.
\end{enumerate}

The rest of this paper is organized as follows: Section~\ref{sec:Methodology} provides the background of utilized neural networks. Section~\ref{sec:Bi-ResNet} explains the proposed Bi-ResNet including the intra links, spatial-temporal convolution bloc, and the network architecture. Then, Section~\ref{sec:experiment} shows experimental results including the test accuracy and the corresponding ablation experiments, and Section~\ref{sec:Conclusion} concludes the paper.

\section{Methodology}\label{sec:Methodology}

\subsection{Convolution Neural Network }

\begin{figure}[!t] 
\centering
\includegraphics[width=0.8\linewidth]{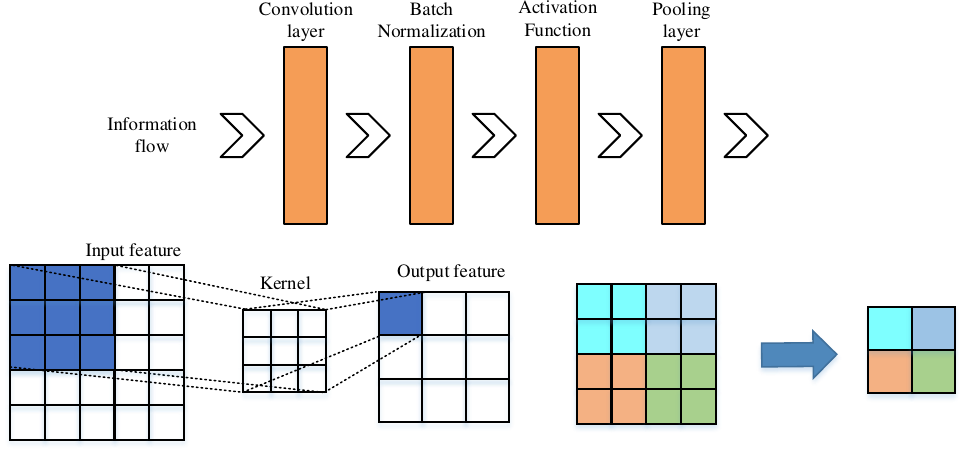}
\caption{The basic components in convolution neural network. The left is the convolution layer and the right is the pooling layer.}
\label{fig:basic structure of CNN}
\vspace{-0.3cm}
\end{figure}

Convolution neural network consists of convolution layers, batch normalization layers, activation functions, and pooling layer, and is optimized by back-propagation algorithm. The basic information flow and two components (convolution layer and pooling layer) are described in Figure~\ref{fig:basic structure of CNN}. By shared-weight architecture of the convolution kernels that slide along input features and provide translation-equivariant responses known as feature maps, CNN can extract different levels of hierarchical features from raw data~\cite{zhang1988shift}. The computation can be concluded by matrix multiplication:
\begin{equation} \label{eq:convolution layer}
\mathbf{Y} = \mathbf{A} \cdot \mathbf{W} +\mathbf{b}
\end{equation}
where $\mathbf{A}$ is sub-input matrix which has the same size as kernel; $\mathbf{W}$ is the kernel which is a matrix as well; $\mathbf{b}$ is the bias; $\mathbf{Y}$ is the calculated feature.  
The activation function provides nonlinearity to the decision function and overall network without affecting the receptive fields of the convolution layers. Rectified linear unit (ReLU) is the most widely utilized activation function which uses the non-saturating function and is shown:
\begin{equation} \label{eq:ReLU}
f(x)=max(0,x)
\end{equation}
where $x$ is the input of ReLU. It effectively removes negative values from an activation map by setting them to zero.
% \cite{romanuke2017appropriate}
% . The ReLU has been found better in training deeper networks~\cite{glorot2011deep}.

The pooling layer is a form of non-linear down-sampling, reducing the spatial size of the representation and helping find the max or mean feature by max or average pooling~\cite{geron2022hands}. The batch normalization layer is created to avoid data distribution shifting as the depth of the neural network increases~\cite{bjorck2018understanding}. It has been proven to be effective in deep modeling.

\subsection{Residual Learning} 

Residual block or shortcut is normally utilized to solve the difficulty introduced by the degeneracy phenomenon in the training process~\cite{he2016deep}. In industrial system fault diagnosis, the processed input is normally a long 1-D complex vibration signal which makes the deeper models more efficient. However, deeper models are difficult to put into practice limiting the improvement of fault diagnosis~\cite{8842598}. From the respective of information theory, as mentioned in Section~\ref{Introduction}, the learned identity mapping can force models to learn high-frequency components from raw data and provide sufficient and necessary information. Ideally, there is no information loss in the forward model since the shortcut enables input data to flow directly to the next layer. 

\begin{figure}[!t] 
\centering
\includegraphics[width=0.8\linewidth]{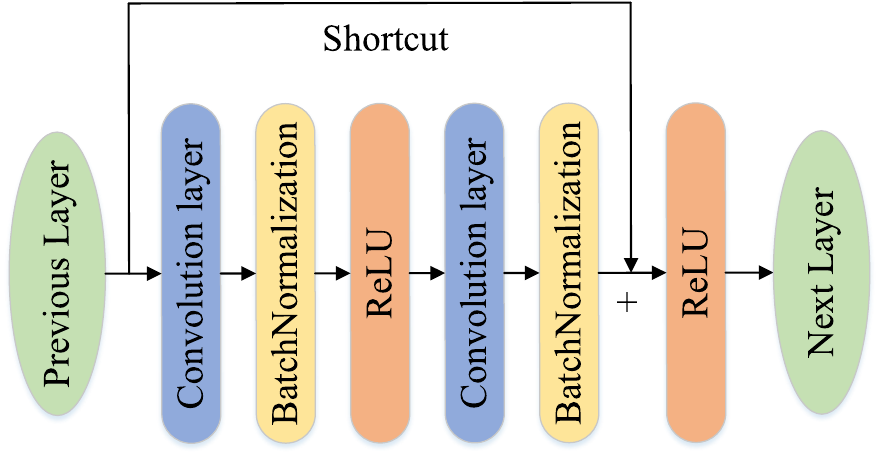}
\caption{The residual block in ResNet18~\cite{he2016deep}.}
\label{fig:residual block}
\vspace{-0.3cm}
\end{figure}

Figure~\ref{fig:residual block} gives an example of residual block which is utilized in ResNet18~\cite{he2016deep}. It explains how the residual block works and information flows, expressed as:
\begin{equation} \label{eq:residual}
 \mathbf{y}^l=F(\mathbf{y}^{l-1},{\mathbf{W}^l,\mathbf{b}^l})+\mathbf{y}^{l-1}
\end{equation}
where $\mathbf{y}^l$ and $\mathbf{y}^{l-1}$ are the output of block $l$ and $l-1$. $\mathbf{W}^l,\mathbf{b}^l$ are the weights and bias of block $l$. $F(\cdot )$ is the residual function.
The input from the previous layer goes through a basic convolution network including the convolution layer, batch normalization, and ReLU, extracting basic features and providing non-linearity. Then, the learned features are again extracted by another convolution layer and data distribution is modified by another batch normalization. Before being input into ReLU, the processed information is added with the original information by shortcut, providing more information for the next layer.  

\section{Bi-residual neural networks} \label{sec:Bi-ResNet}
%小标题可以再想一想 跟上一节合起来？

As the scaling law theory is widely confessed and applied in computer science, large models seem to become the answer to all problems. Nevertheless, in contrast to large models, industrial models are more focused on solving a certain type of problem rather than all problems, saving computing resources. The Bi-ResNet is proposed in this section to discuss how small models can also do big things as long as the architecture is well-designed.

\subsection{Intra-linked Layer}

\begin{figure}[!t] 
\centering
\includegraphics[width=0.95\linewidth]{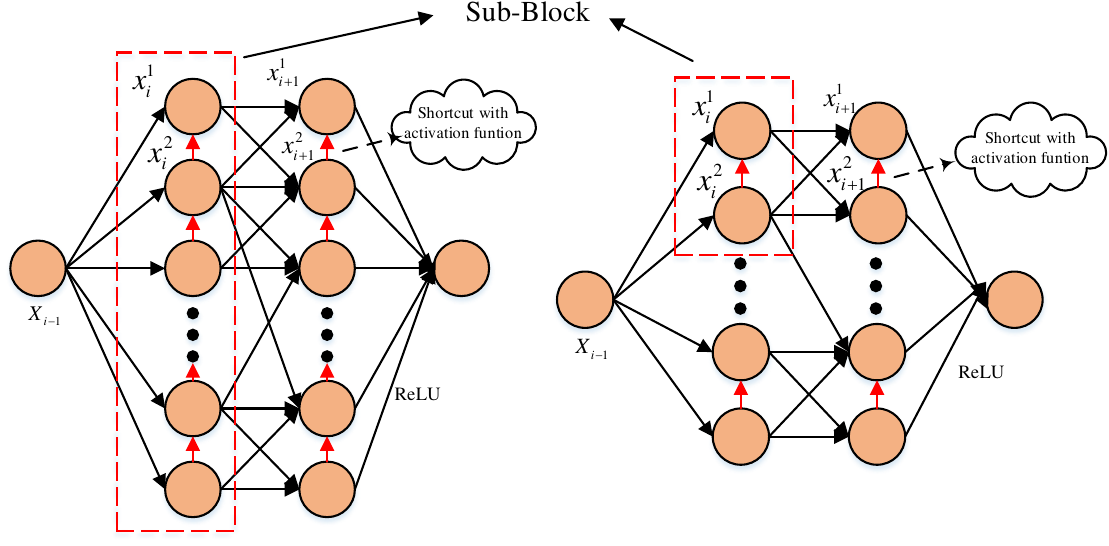}
\caption{The schematic diagram of the intra-linked layer. The left is a model with sequential intra-link layers. The right is a model whose neurons in each layer are connected two by two.} %%搞个超参出来 n
\label{fig:intra-linked layer}
\vspace{-0.3cm}
\end{figure}

Inspired by \cite{chen2021amplitude}, we notice that from the perspective of information theory, the inner neurons can be connected by a shortcut achieving a similar architecture as the residual block named horizontal residual block or intra-links.
The block can be recognized as a built-in high-dimensional signal extractor without adding additional parameters. In this block, there is a hyper-parameter $n$ that determines how many neurons are connected (the mode of neuron connection). Assuming that there are $m$ neurons in the intra-link block.
Figure~\ref{fig:intra-linked layer} gives two examples of models with intra-link layers. The left one is the maximum mode where $n=m-1$ and the right one is minimum mode where $n=1,m=2$. For the conventional models such as fully connected models or CNN, the $n=0$.
% The difference between the two models depends on how many neurons in the same layer are connected. 
From the perspective of network architecture, the intra-link can be recognized as the height of networks which may improve a network’s representation capability.

Assuming that for an $\mathbb{R}^{w_0}\rightarrow \mathbb{R}$ ReLU deep neural neural network with widths $w_1, w_2, \dots, w_k$ of $k$ hidden layers, the input of the model can be expressed as $ \mathbf{x_0}=[x_{0}^{1}, x_{0}^{2}, \dots, x_{0}^{w_0}]\in \mathbb{R}^{w_0}$ and the output vector of i-th layer is $\mathbf{x_i}=[x_{i}^{1}, x_{i}^{2}, \dots, x_{i}^{w_i}]\in \mathbb{R}^{w_i}, i=1, 2, \dots, k$.
For each intra-link block in intra-link layers, there are two different computation modes. For the neurons that accept outputs of other neurons in the same layer as inputs, the pre-activation of the j-th neuron in the i-th layer and the corresponding neuron is given by
\begin{equation} \label{eq:intra link left}
\begin{split}
&g_i^j=\left \langle \mathbf{a}_i^j,\mathbf{x}_{i-1} \right \rangle +{b}_i^j +\sigma(g_i^{j+1}) \\
&x_i^j = \sigma(g_i^{j})
\end{split}
\end{equation}
where $\mathbf{a}_i^j$ and ${b}_i^j $ are separately the weight and bias vector of j-th neurons. $\sigma(\cdot)$ is the activation function. The rest neurons with no additional inputs follow the conventional fully connected networks. The pre-activation of the j-th neuron in the i-th layer and the corresponding neuron are given by
\begin{equation} \label{eq:intra link right}
\begin{split}
&g_i^j=\left \langle \mathbf{a}_i^j,\mathbf{x}_{i-1} \right \rangle +{b}_i^j \\
&x_i^j = \sigma(g_i^{j})
\end{split}
\end{equation}

In Figure~\ref{fig:intra-linked layer}, there is only one neuron in one layer that follows Eq~\ref{eq:intra link right} in the left model. For the right model, half of the neurons follow the conventional fully connected networks (Eq~\ref{eq:intra link right}) and the rest follow the inner connection law (Eq~\ref{eq:intra link left}). Since there are no additional parameters added to networks, the complexity of the model is not increased. The computation complexity depends on the modes of inner connection. 
% For the model to be deployed, the left connection mode is utilized in this paper.

\subsection{Embedded Spatial-Temporal Convolution Block}
The fault signals contains the spatio-temporal features which include not only temporal variation characteristics but also spatial distribution. The accurate feature extraction can help identify and locate faults and improve the accuracy and efficiency of diagnosis. Therefore, we design an embedded spatial-temporal convolution block to capture the pattern of fault signals in Bi-ResNet as a part of the intra-linked layer.

\begin{figure}[htbp] 
\centering
\includegraphics[width=0.8\linewidth]{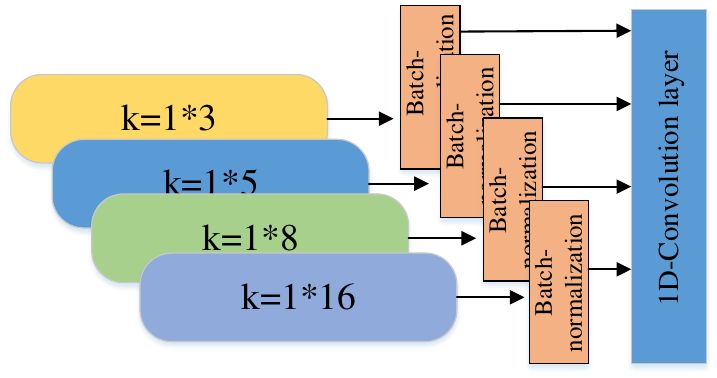}
\caption{The embedded spatial-temporal convolution block.}
\label{fig:Spatial-temporal convolution block}
\vspace{-0.3cm}
\end{figure}

Figure~\ref{fig:Spatial-temporal convolution block} shows the designed embedded multi-scale embedded spatial-temporal convolution block which consists of four parallel 1D-convolution layers with different kernel sizes, four batch-normalization layers, and one global 1D-convolution layer. Specifically, the inputs from the previous layer are first extracted to get temporal features at multiple scales. Subsequently, the features go through a batch-normalization layer to ensure uniform distributions. Then, we use a global 1D-convolution layer to extract spatial information. This block is embedded into the intra-link layers as a feature extractor and the extracted features will be further processed to extract high-frequency spatio-temporal features.

\subsection{Bi-ResNet Structure}

\begin{figure*}[!t] 
\centering
\includegraphics[width=0.95\linewidth]{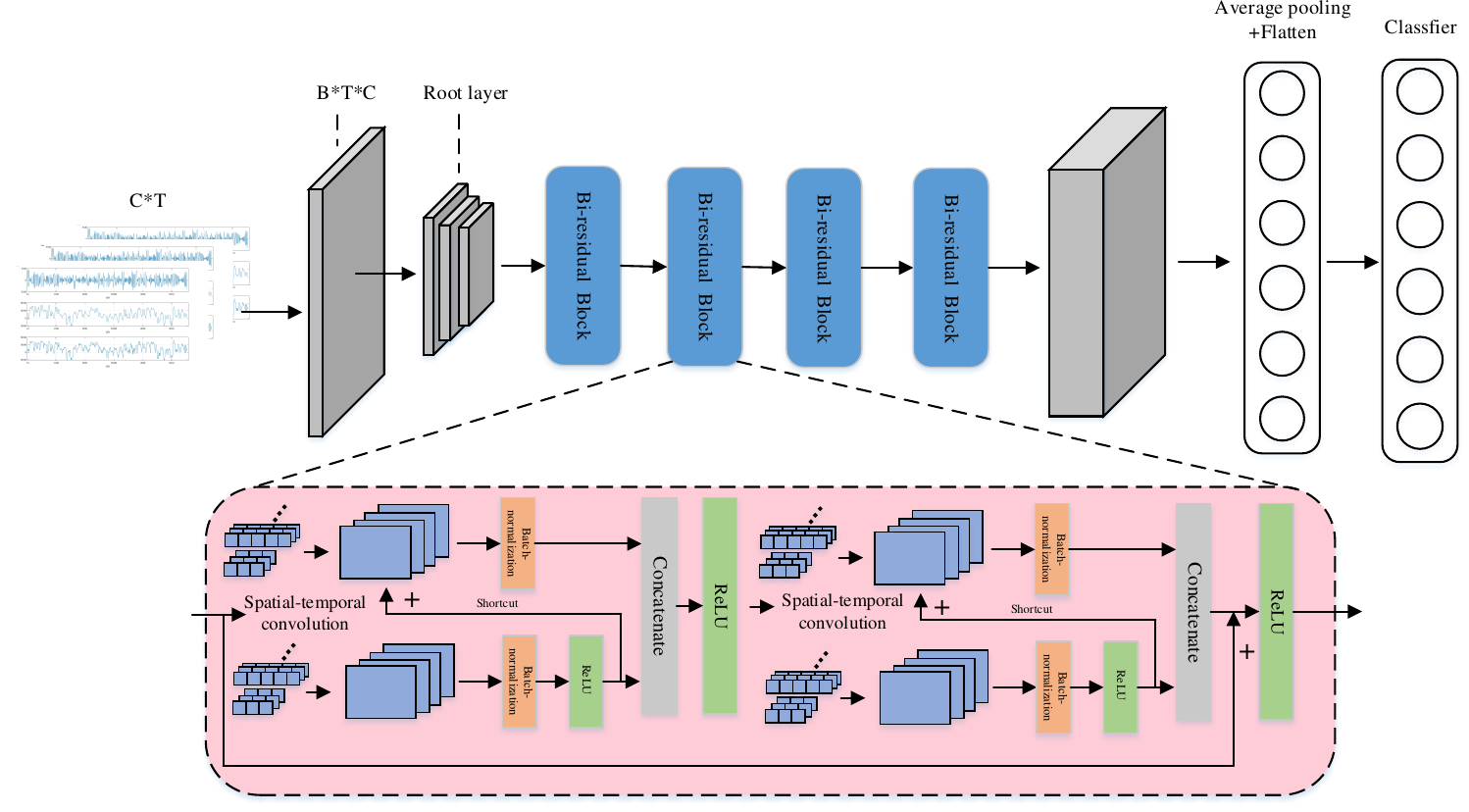}
\caption{The structure of Bi-ResNet.}
\label{fig:The structure of Bi-ResNet}
\vspace{-0.3cm}
\end{figure*}

The new designed architecture of Bi-ResNet is shown in Figure~\ref{fig:The structure of Bi-ResNet}.
% inspired by ResNet18 which has demonstrated its effectiveness in image processing and language models. Figure~\ref{fig:The structure of Bi-ResNet} shows the whole structure of Bi-ResNet. 
The input signals are reconstructed into 3 dimensions adapting to the 1D convolution mode of the root layer which is composed of one 1D convolution layer, bath-normalization layer, and activation layer. It is utilized to extract features from data on a large scale, preparing for subsequent high-frequency extraction. Then, the extracted signals are input into four basic blocks which are residual modules consisting of embedded spatial-temporal convolution blocks with intra-link followed by batch-normalization layers and ReLU.  At last, all inputs go through the average pooling layer and are flattened for classification.

The intra-link mode is chosen based on the complexity of the datasets. As described in Figure~\ref{fig:The structure of Bi-ResNet} where $n=1$, the neurons at the same layer are connected two by two. From the respective of information theory, half neurons accept the non-linear information from the rest neurons at the same layer, and then the information from all neurons is concatenated as the input of the next layer. Part of the information is reused twice which can be considered as the inner residual connection at one layer.

% The basic block is also inspired by the 

\section{Experimental Results} \label{sec:experiment}

To further access the efficacy of the proposed model, we evaluate Bi-ResNet against a wide range of deep learning models, including deep 1-D convolution neural network(CNN-1)~\cite{10472098}, 1D-CNN+LSTM(CNN-2)~\cite{kaya2024efficient}, 2DCNN-GRU(CNN-3)~\cite{li2023fault}, ResNet18~\cite{he2016deep} and two golden baselines(CNN-4 and LSTM)~\cite{sun2023public}, since these models have demonstrated their out-performance in synchronous motor electrical faults diagnosis. We conduct an experiment about synchronous motor electrical faults for this, generating numerous failure data.
CNN-1 is a traditional 1D-CNN network. CNN-2 has both 1D convolution layers and LSTM layers to extract the potential time-related features. CNN-3 uses 2D-CNN and GRU instead of 1D-CNN and LSTM to extract features. These three models do not have the residual learning mechanism in modeling. In contrast, ResNet18 is the first CNN-based model to use residual networks and achieves good performance in many aspects. 
% The other two models are the baseline models proposed together with the dataset. 
Table~\ref{Table: The PRM of all models} shows the number of parameters of the main models, explaining the computation burden of each model.
We conduct these comparisons across motor fault datasets with multi-resolution and multi-level noise. 

\begin{table}[!t]
\centering\footnotesize
\caption{The PRM of main models. \#PRM denotes the number of parameters.}
\label{Table: The PRM of all models}
\renewcommand{\arraystretch}{1.3}
\scalebox{1}{\begin{tabular}{ l| l l l l l}
\toprule
Models & CNN-1 & CNN-2 & CNN-3 & Bi-ResNet & ResNet18  \\
\hline
\#PRM & 1.01M  & 1.12M & 1.17M  & 1.05M & 11.2M  \\
\bottomrule 
\end{tabular}}
\vspace{-0.3cm}
\end{table}

%%%%%% 没有高频的数据对比， 多个下采样间隔的对比 多个noise的对比 是不是真的找到高频数据，怎么体现？ 内连接多个对比  内连接放在其他模型中 

\subsection{Experiment Description} %  实验图像 低分辨率-下采样
\begin{figure}[!t]
\centering
\includegraphics[width=0.95\linewidth]{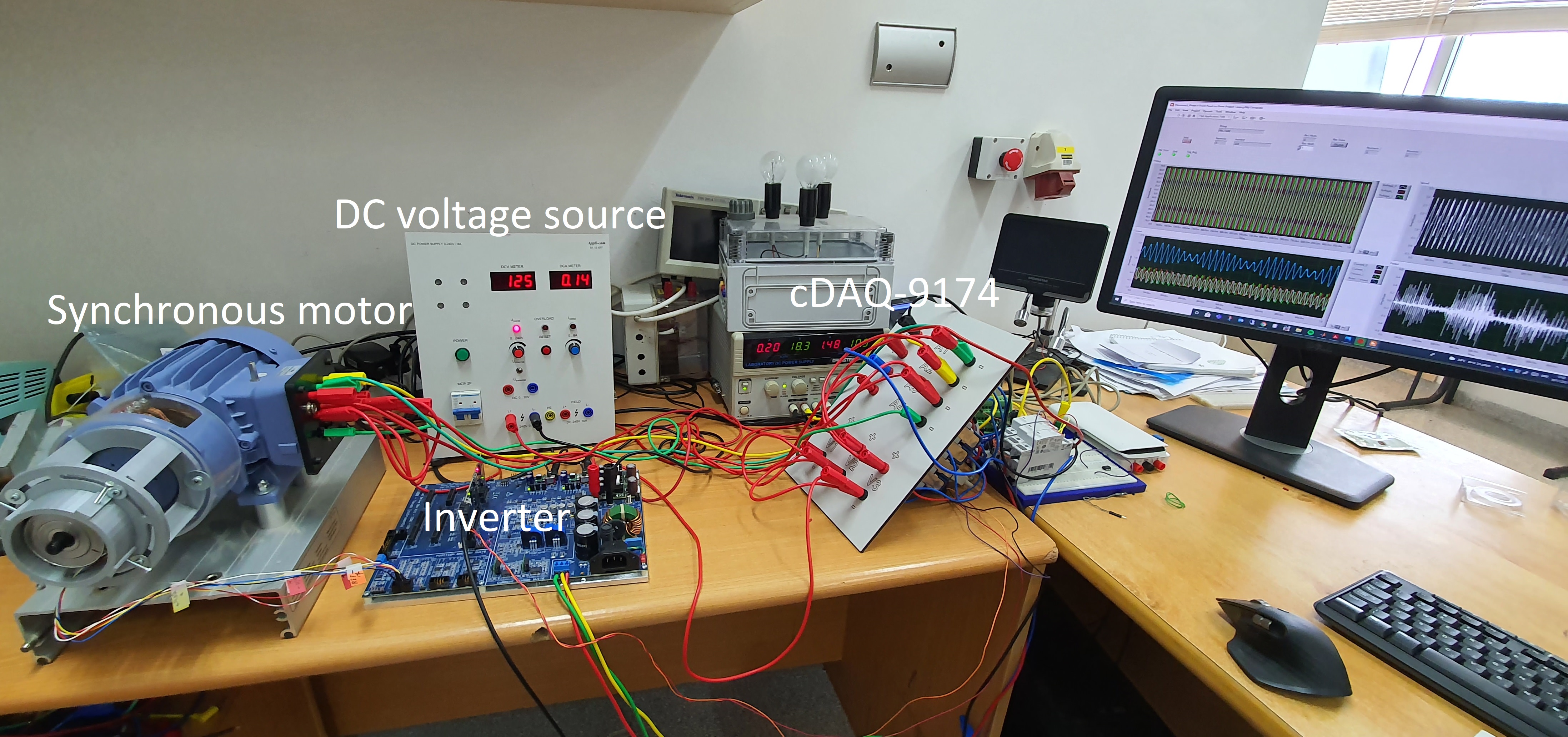}
\caption{Experimental setup }%~\cite{sun2023public}.}
\label{fig:exp_setup}
\vspace{-0.3cm}
\end{figure}

The experimental dataset is collected for five types of motor faults listed in Table~\ref{tab:Classification labels}, using a synchronous motor, a DC voltage source (125V), an inverter, and a cDAQ-9174 which is shown in Figure~\ref{fig:exp_setup}. 
The DC voltage source supplies a three-phase, two-level inverter, which is realized through the TMDXIDDK379D development kit from Texas Instruments. This kit features an integrated servo drive with a complete power stage, enabling it to drive a three-phase synchronous motor while accommodating position feedback, current sensing, and control mechanisms. The parameters analyzed within this system—specifically the stator phase currents and voltages, along with rotor current and speed—are captured using the NI cDAQ-9174 portable sensor measurement system. The cDAQ-9174 is equipped with current and voltage sensors, managing the timing, synchronization, and data transfer between NI C Series I/O modules and an external computer.
The mathematical model of the synchronous motor in the dq0 reference frame, for both steady and transient states, is given by
\begin{equation} \label{eq:motor fault model}
\scalemath{0.8}{
\begin{split}
&\frac{d\theta }{dt} = w \\
&\frac{dw }{dt} = \sigma (3P_{ref}-\frac{1}{D}(w_m-w_s)+\frac{3\beta w_s-6L_{ff} L_q w_s}{2\beta L_q}\lambda_d \lambda_q +\frac{3L_{af}w_s}{\beta}\lambda_q \lambda_f )\\
&\frac{d\lambda_d }{dt}=-\frac{2R_{a}L_{ff}}{\beta}\lambda_{d}+w_{m}\lambda_{q}+\frac{2R_{a}L_{af}}{\beta}\lambda_{f}+\upsilon_{d}\\
&\frac{d\lambda_{q}}{dt}=-w_{m}\lambda_{d}-\frac{R_a}{L_q}\lambda_{q}+\upsilon_{q}\\
&\frac{d\lambda_{0}}{dt}=-\frac{R_a}{L_0}\lambda_{q}+\upsilon_{0}   \\
&\frac{d\lambda_{f}}{dt}=\frac{3R_{f}L_{af}}{\beta}\lambda_{d}-\frac{2R_{f}L_{d}}{\beta }\lambda _{f}+\upsilon_{f}  \\
\end{split}
}
\end{equation}
where $w_s=2\pi f$ is the grid frequency, $R_a$ is armature resistance, $R_f$ denotes field winding resistance,$\lambda_{f}$ stands for field winding flux linkage, $\lambda_{d}, \lambda_{q}, \lambda_{0}$ are dq0 transformations of the stator flux linkages,
$\upsilon_{d}, \upsilon_{q}, \upsilon_{0}$ are dq0 transformations of stator voltages, $\upsilon_{f}$ is field winding voltage, $L_{ff}$ is self-inductance of the field winding, $L_{af}$ is stator-rotor mutual inductance per phase, and $L_{d}, L_{q}, L_{0}$ are the direct-axis, quadrature-axis and zero-sequence synchronous inductance, respectively. The coefficients $\sigma$ and $\beta$ are obtained from:
\begin{equation} \label{eq:motor fault model}
\scalemath{1}{
\begin{split}
&\sigma=(\frac{P}{2})^{2}\frac{1}{Jw_s}   \\
&\beta=2L_{d}L_{f}f-3L_{af}^{2}\\
\end{split}
}
\end{equation}

In the experiments, each fault has hundreds of experiments and each experiment contains 10,000 samples and 10 features, including three-phase voltages, three-phase currents, motor speed, motor current, and time point of failure. The duration of each experiment is one second.
% The dataset has been pre-processed in~\cite{sun2023public} for simple fault diagnosis.
Due to the sampling time of 1ms, which means the high resolution of the dataset, we performed low resolution downsampling on the dataset such as 2ms, 5ms, 10ms, and 20ms for the dataset diversity.

% The dataset is a public experimental dataset.  Figure~\ref{fig:exp_setup} shows the experiment setup including a synchronous motor, a DC voltage source, an inverter, and a cDAQ-9174 whose details are introduced in~\cite{sun2023public}. The generated dataset concludes five motor faults which are shown Table~\ref{tab:Classification labels}. Each fault has hundreds of experiments, and each experiment contains 10,000 samples and 10 features including three-phase voltages, three-phase currents, motor speed, motor current, and time point of failure. The sample time is 1 ms which means the high-resolution of the dataset. Hence, we down-sample the raw data for low resolution.

\begin{table}[!t]
\centering\small
\caption{Classification labels of synchronous motor electrical faults.}
\label{tab:Classification labels}
\renewcommand{\arraystretch}{1.3}
\begin{tabular}{l l}
\toprule
Class & Explanation \\ 
\hline
REVD & Rotor excitation voltage disconnection \\
OP & Opened phase \\
VREC & Variation of rotor excitation current \\
2PSC & Two phases short circuit \\
1PSC & One phase-to-neutral short circuit \\
NF & No fault \\
\bottomrule 
\end{tabular}
\vspace{-0.3cm}
\end{table}

\begin{figure}[!t]
\centering
\subfloat[Voltage Example]{
\includegraphics[width=0.9\linewidth]{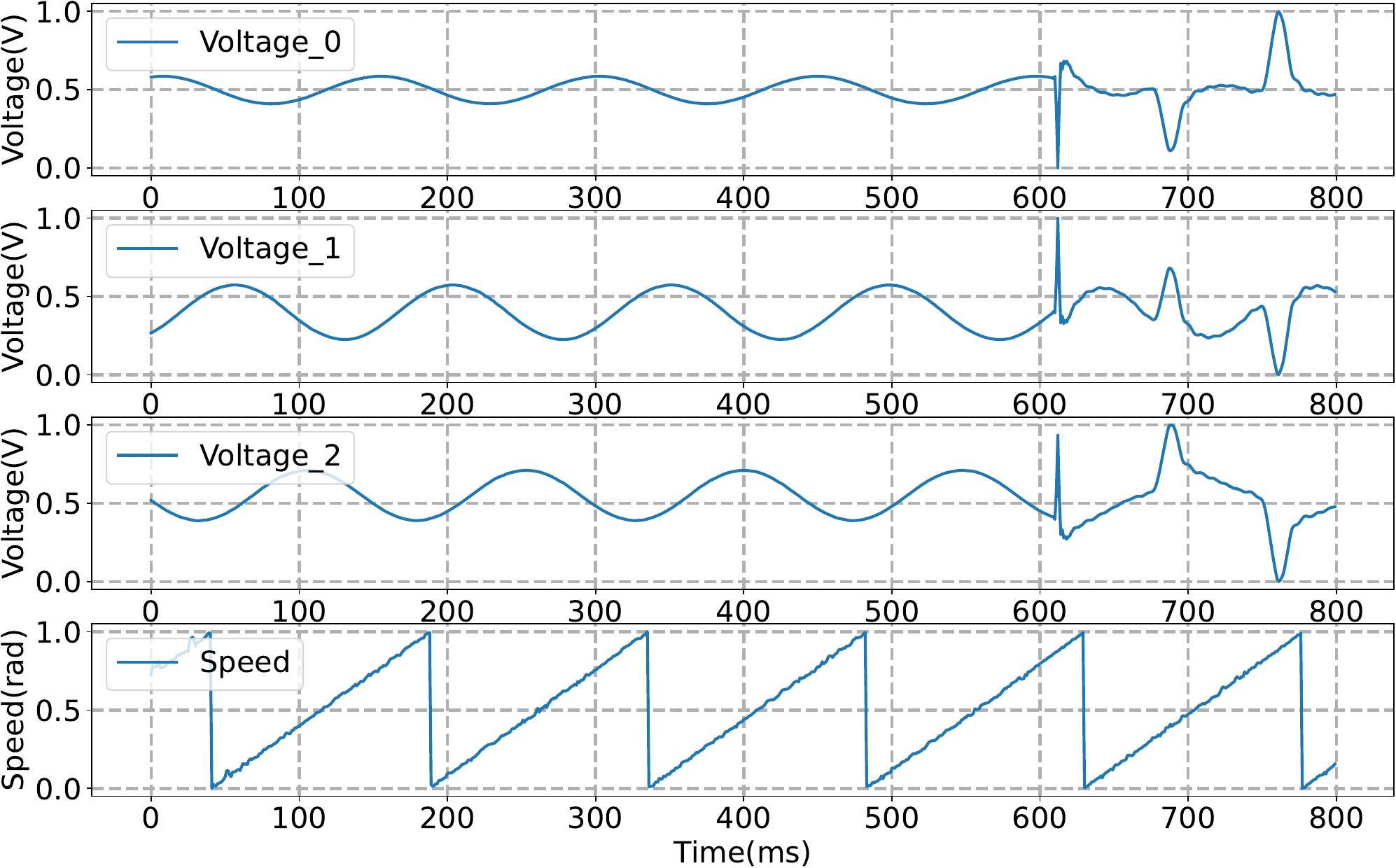}
\label{fig:Data analysis voltage}
} \\
\subfloat[Current Example]{
\includegraphics[width=0.9\linewidth]{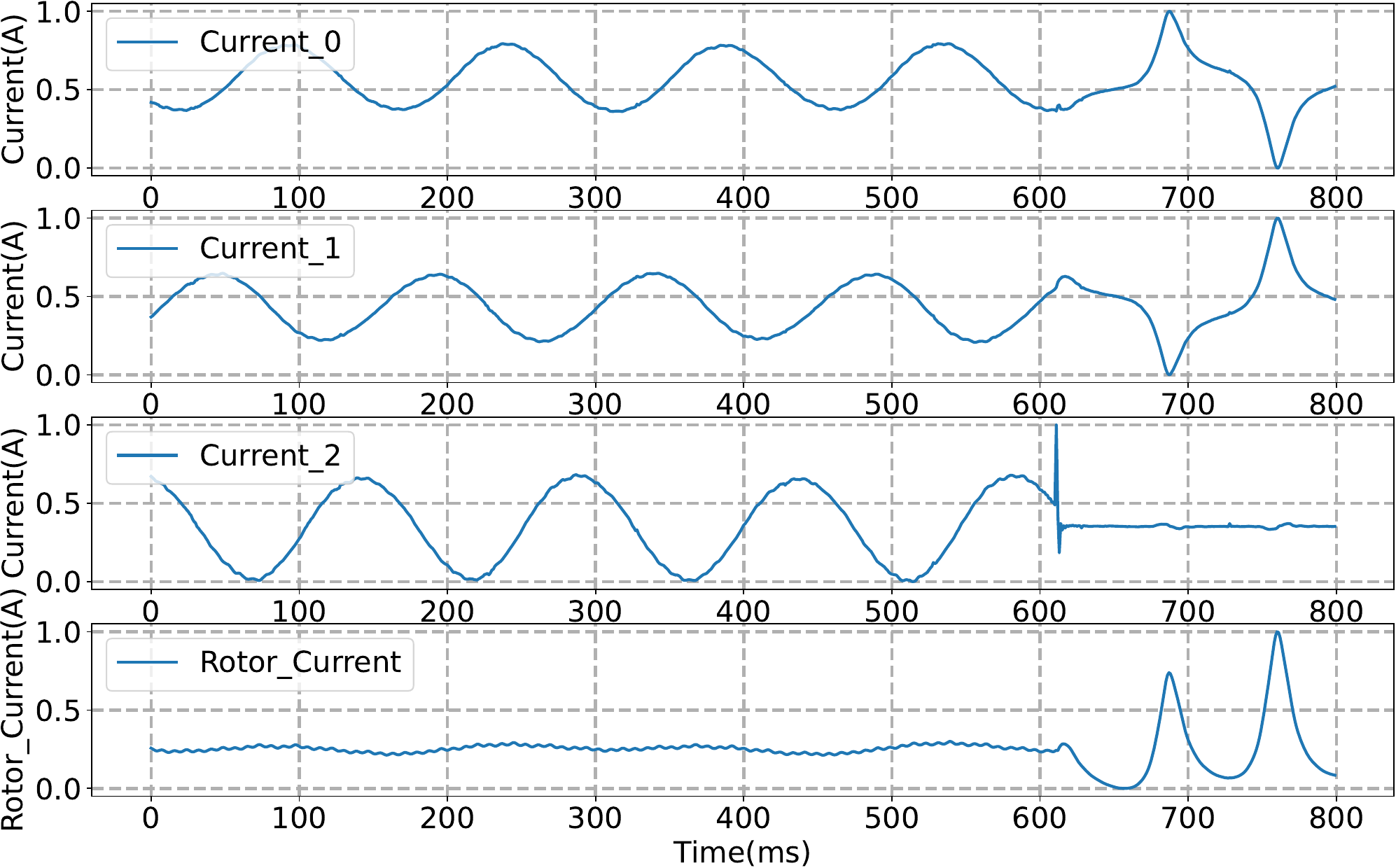}
\label{fig:Data analysis current}
}\\
\caption{Data analysis of experimental dataset. }
\label{fig:Data analysis}
\vspace{-0.3cm}
\end{figure}

% Figure~\ref{fig:Data analysis} takes opened phase fault as an example showing the normalized details of datasets (800 samples). Compared with the normal signals, the distortion of voltage and current belongs to high-frequency components and needs to be extracted. In addition, there is little noise in the dataset which is opposite to the actual situation. Thus, additional noise is added to explore the potential of Bi-ResNet.
Figure~\ref{fig:Data analysis} takes opened phase fault as an example showing the normalized details of datasets (800 samples). Compared with the normal signals, the distortion of voltage and current belongs to high-frequency components and needs to be extracted. In addition, since the laboratory experiments are relatively simple, in order to simulate more practical and complex engineering problems, we add additional noise with different levels of signal-to-noise ratio (SNR) to the dataset.

\subsection{Model Setup} %
For each experiment, we train the models with a batch size of 64 using the ‘Adam’ optimizer
% ~\cite{kingma2017adammethodstochasticoptimization} 
with an initialized learning rate of $0.01$. The learning rate decays with a factor of 0.5 if the accuracy at 10 consecutive epochs is unchanged. The epoch is set to 100. To constrain the weights of models, the L2 regularization is used in each experiment. The dataset is split into training, validation and test sets with a ratio of 0.8:0.1:0.1. The experiments are implemented in 'Tensorflow' using a CPU Intel i7-11800H processor at 2.3Hz and a GPU NVIDIA T600. The utilized Bi-ResNet architecture is shown in Table~\ref{tab:Bi-ResNet Architecture}. For the model to be deployed, the minimum mode is utilized in this paper ($n=1$).

\begin{table}[!t]
 \centering
 \caption{Bi-ResNet Architecture}
 \label{tab:Bi-ResNet Architecture}
 \renewcommand{\arraystretch}{1.3}
 \begin{tabular}{c|c|c}
 \toprule
  Layer & Parameters & Activation \\
 \hline
 {Convolution ($3\times1$)} & Filter size = 32, Stride = 1  &  ReLU \\
 \hline
 Batch-normalization (BN) & momentum=0.99, epsilon=0.001 & --\\
 \hline
 Bi-residual block (*2) & Filter size = 32, Stride = 1 & ReLU\\
 \hline
 Bi-residual block (*2) & Filter size = 64, Stride = 1 & ReLU\\
 \hline
 Bi-residual block (*2) & Filter size = 128, Stride = 1 & ReLU\\
 \hline 
 Bi-residual block (*2) & Filter size = 256, Stride = 1 & ReLU\\
 \hline
 \multicolumn{3}{c}{Global MaxPooling} \\
 \hline 
\multicolumn{3}{c}{Flatten layer} \\
 \hline
 Fully connected & Size = 6 & Softmax \\
 \bottomrule 
 \end{tabular}
 \vspace{-0.3cm}
 \end{table}

\subsection{Analytical Experiments} %多保证率
The analytical experiment explores the efficiency of Bi-ResNet at different data resolutions. By choosing different sample times, the dataset can be down-sampled with various resolutions.

Table~\ref{Table: Analytical Experiments different sample time} summarizes the test accuracy of all mentioned models with different sample times using synchronous motor electrical faults with/without noise. In noise-free data, CNN-based models all have comparative performance. When we use a high-resolution dataset (sample times are 1ms and 2 ms), almost all the models can achieve 100\% test accuracy. As the sampling time increases, the advantages of the residual learning gradually emerge. Bi-ResNet and ResNet18 can both guarantee about 99\% test accuracy, far exceeding other models.

\begin{table*}[!t]
\centering\footnotesize
\caption{The test accuracy of motor fault diagnosis with different sample times using a dataset with/without noise.}
\label{Table: Analytical Experiments different sample time}
\renewcommand{\arraystretch}{1.3}
\scalebox{1}{\begin{tabular}{c| c| l l l l l l l}
\toprule
Noise & Sample Time & CNN-1 &CNN-2 & CNN-3 & Bi-ResNet & ResNet18 & CNN-4 & LSTM  \\
\hline
\multirow{5}{*}{Dataset without noise} & 1 ms & 100\% &100\% & 100\% &100\% & 100\% & 100\%& 85.36\%   \\
 & 2 ms & 100\% &100\% & 100\% &100\% & 100\% & 99.76\%& 85.31\%\\
 & 5 ms & 99.68\% &99.94\% & 100\% &100\% & 100\% & 97.32\%& 83.36\% \\
 & 10 ms & 97.96\% &98.46\% & 99.43\% &100\% & 100\% & 94.26\%& 77.89\% \\
 & 20 ms & 95.56\% &95.35\% & 94.54\% &99.86\% & 98.73\% & 92.69\%& 74.34\%  \\
\hline
\multirow{5}{*}{Dataset with noise} & 1 ms & 100\% &100\% & 100\%&100\% &100\% &100\% &  85.10\%  \\
 & 2 ms & 99.96\% &100\% & 100\%&100\% &100\% &98.64\% &  85.10\%  \\
 & 5 ms & 96.54\% &98.12\% & 98.54\% & 100\% &100\% &95.12\% &  82.28\%  \\
 & 10 ms & 95.78\% &94.57\% & 96.86\% & 99.14\% &97.98\% &92.46\% &  73.67\%  \\
 & 20 ms &92.11\%  & 91.86\% & 92.68\% & 94.82\% & 94.17\% & 90.20\% & 72.32\% \\
\bottomrule 
\end{tabular}}
\vspace{-0.3cm}
\end{table*}

When additional noise is added to the dataset, the potential of Bi-ResNet is highlighted especially at low-resolution datasets (sample times are 10ms and 20 ms). At high-resolution datasets, models can all perform as well as at noise-free datasets. However, when the sample time becomes 10ms, the test accuracy drops significantly except Bi-ResNet. 
Bi-ResNet can extract necessary high-frequency components from raw data and noise signals for motor fault diagnosis. Residual learning is considered to be an efficient high-frequency signal extractor. At the same time, the inner connection is equivalent to a built-in sub-signal extractor in the extractor, which helps the model to perform more accurate classification without increasing parameters. Meanwhile, since no more parameters are added to models, the over-fitting can also be avoided. The above analytical experiments demonstrate the efficacy of Bi-ResNet.

\subsection{Comparative Experiments} %不同的噪声， 不同模型的内连接
To further test the robustness of Bi-ResNet, noise with different levels of signal-to-noise ratio is added to the dataset for comparison. Meanwhile, models including 1D-CNN(CNN-1)~\cite{10472098}, 1D-CNN-LSTM(CNN-2)~\cite{kaya2024efficient} and 2DCNN-GRU(CNN-3)~\cite{li2023fault} are modified by intra links to figure out that whether intra links can be applied in different network architecture achieving better performance.

\subsubsection{Performance under different SNR} Noise is unavoidable in the real-life dataset which is different from the experimental dataset. The level of noise will directly affect the classification performance of models. We add additional noise to our dataset, setting the SNR to -5, -3, -1, 1, 3, 5 respectively. The sample time is 10ms to ensure low resolution and model performance distinction.  

Table~\ref{Table: experiments with different SNR} compares the test accuracy of models on synchronous motor electrical fault dataset across multiple SRNs. The Bi-ResNet achieves the best performance among all the experiments. When $SNR>0$, models can maintain similar performance except for two simple baselines. However, as the proportion of white noise increases, the advantages of residual learning and intra-links emerge. Bi-ResNet takes the lead by the largest margin, i.e., surpassing the second place ResNet18 by over 1\% and the third place by about 3\%. When $SNR=-5$, Bi-ResNet can still keep first place achieving 94.15\% test accuracy, in spite of that the white noise masks some high-frequency fault components.

\begin{table}[!t]
\centering\footnotesize
\caption{The test accuracy of motor fault diagnosis with different SNR.}
\label{Table: experiments with different SNR}
\renewcommand{\arraystretch}{1.3}
\scalebox{0.95}{\begin{tabular}{ l| l l l l l l}
\toprule
Models & -5 & -3& -1 & 1 & 3 &  5   \\
\hline
CNN-1 & 91.65\% &93.55\% & 95.78\% &96.43\% & 96.98\% & 97.56\%    \\
CNN-2  & 90.89\% &92.66\% & 94.57\% & 96.75\% & 97.67\% & 98.37\% \\
CNN-3 & 92.34\% & 95.36\% & 96.86\% & 97.74\% & 98.77\% & 99.41\% \\
Bi-ResNet & 94.15\% &98.28\% & 99.14\% &99.59\% & 99.98\% & 100\% \\
ResNet18 & 93.87\% &97.12\%  & 97.98\% &98.86\% &99.24\% & 100\%  \\
CNN-4  & 88.76\% &89.96\% & 92.46\% & 92.98\% & 93.49\% & 94.14\%  \\
LSTM & 70.35\% &71.69\% & 73.67\% &75.05\% & 75.88\% & 76.35\% \\
\bottomrule 
\end{tabular}}
\vspace{-0.3cm}
\end{table}

\begin{figure}[!t]
\centering
\subfloat[Occlusion experiment example (1PSC)]{
\includegraphics[width=0.9\linewidth]{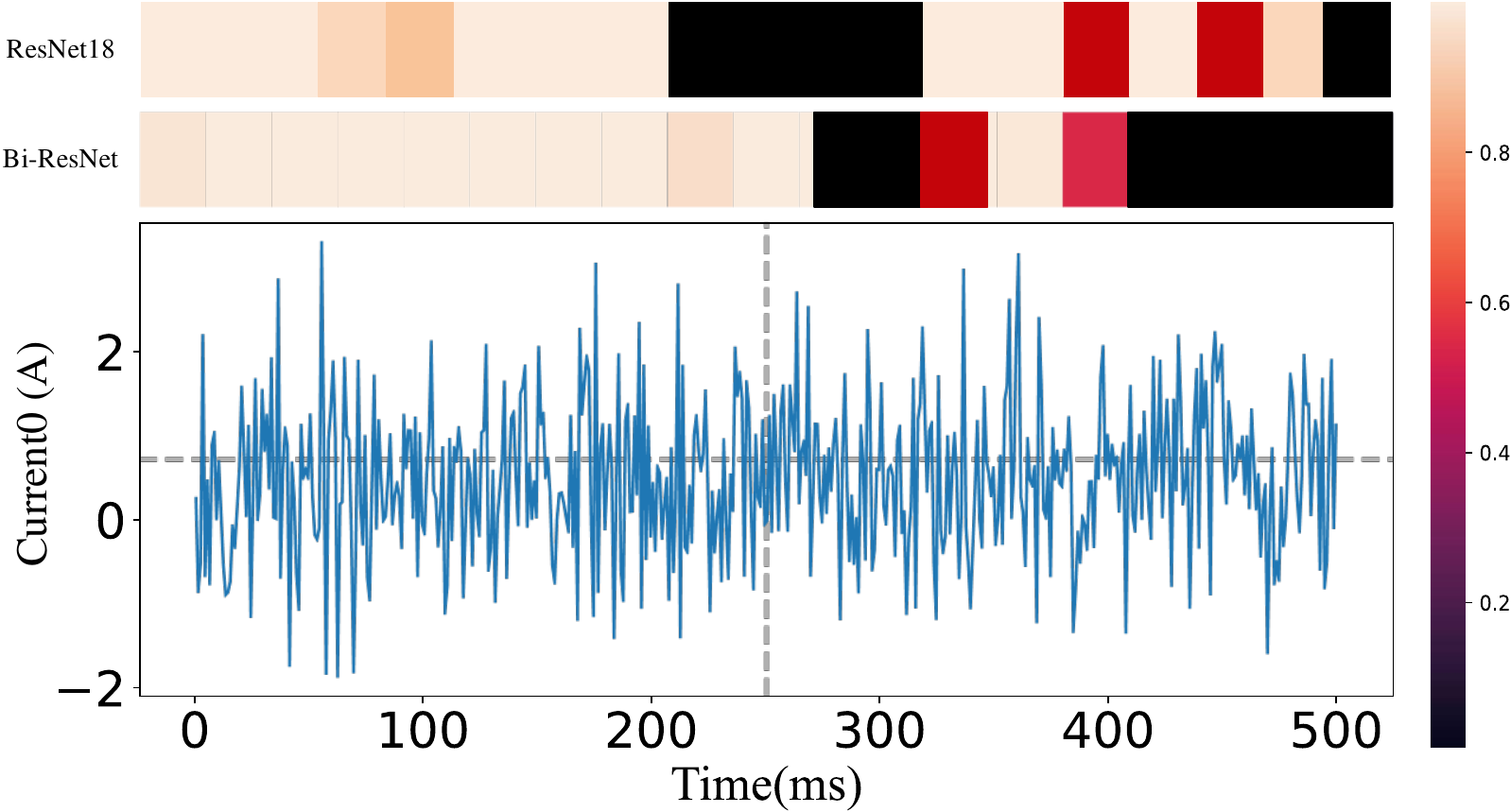}
\label{fig:Occlusion example}
} \\
\subfloat[Corresponding raw current]{
\includegraphics[width=0.9\linewidth]{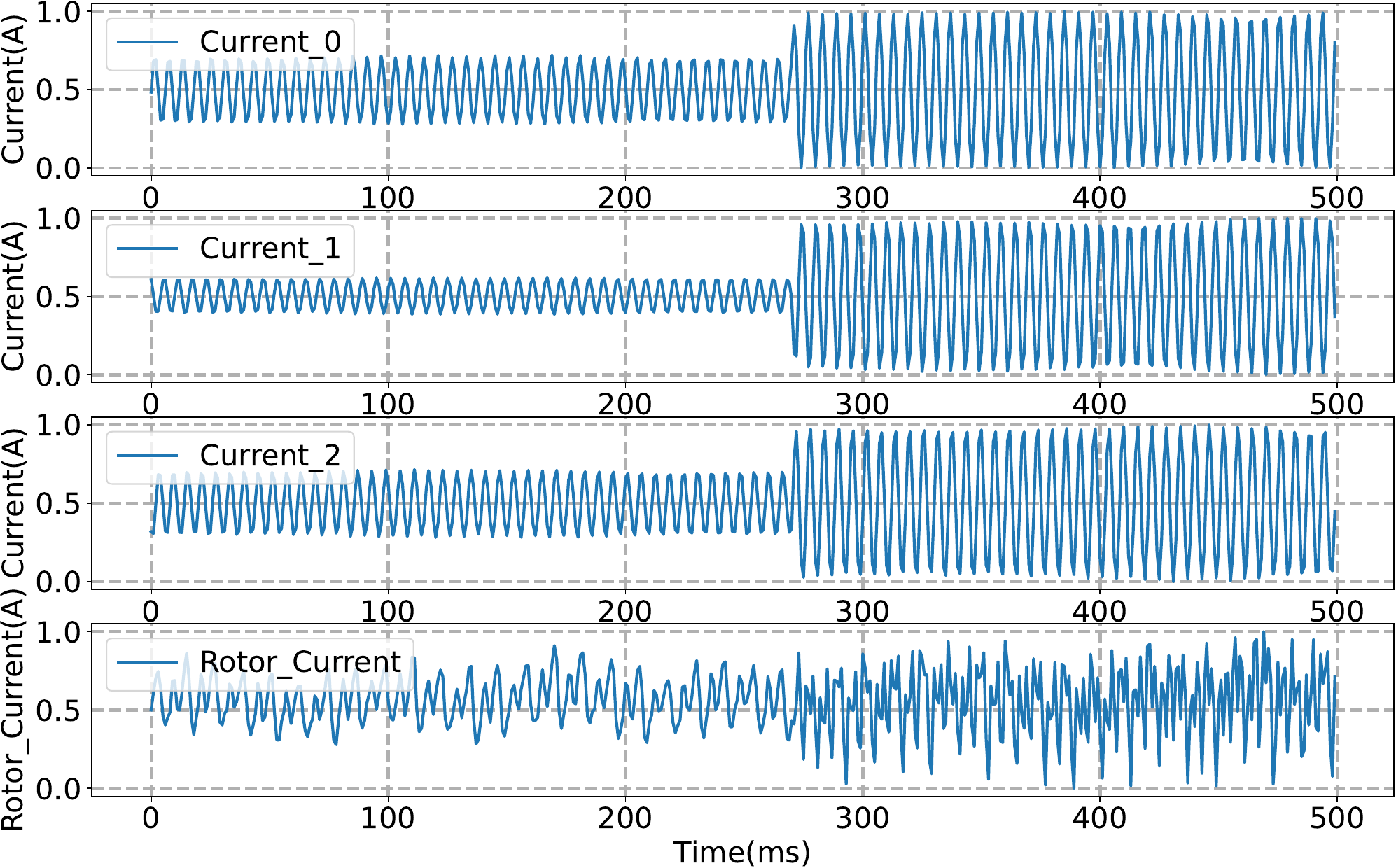}
\label{fig:occlusion experiments current}
}\\
\caption{The extracted features by Bi-ResNet.}
\label{fig:The extracted features by Bi-ResNet}
\vspace{-0.3cm}
\end{figure}

To further evaluate the effectiveness and underline the fault location ability of Bi-ResNet, we conduct occlusion experiments at $SNR=-1$ showing the extracted features by Bi-ResNet and ResNet18, since their test accuracy is better than other models. We take the current of 1PSC as an example, as shown in Figure~\ref{fig:The extracted features by Bi-ResNet}. The occluding sizes and strides were set to 50 and 25 respectively. The occluded pixels were all replaced by zeros. As deposited in the raw data, the fault happens at around 270ms and the three-phase current amplitude has a sudden change. The fluctuation of the rotor current becomes obvious. The occlusion experiment example describes extracted features from the input $current0$ with additional white noise. Despite the noise covering the original signals, the Bi-ResNet can still locate the fault precisely, benefiting the intra-links, the built-in high-frequency component extractor. In contrast, compared with Bi-ResNet, although ResNet18 has competitive test accuracy, it is still inferior in high-frequency component localization. Part of normal signals are considered as the key features of motor fault diagnosis and more fault signals are ignored.

\subsubsection{Performance of different architecture with intra links} To test the scalability of intra links in other deep learning architecture, we add inner connections in 1D-CNN, 1D-CNN-LSTM, and 2DCNN-GRU models, figuring out whether they can improve the test accuracy. The SNR of the utilized dataset is -1 and the sample time is 10ms. Since two baselines have the worst performance in the above experiments, they are not considered in this experiment.

Table~\ref{Table: The scalability of intra links} compares the test accuracy difference between the models with and without intra links. There is an increase in accuracy among all the models, indicating the effectiveness of intra-links. Compared with ResNet18, the test accuracy of CNN-3 is competitive, demonstrating the similarity between inner links and residual learning. Therefore, the intra-links can also be regarded as a kind of residual learning.

\begin{table}[!t]
\centering\footnotesize
\caption{The scalability of intra links.}
\label{Table: The scalability of intra links}
\renewcommand{\arraystretch}{1.3}
\scalebox{1}{\begin{tabular}{ l| l l }
\toprule
Models & Intra links & No intra links   \\
\hline
CNN-1 & 96.98\%  &  95.78\%    \\
CNN-2 & 96.54\%  &   94.57\%     \\
CNN-3 & 97.32\%  &   96.86\%     \\
\bottomrule 
\end{tabular}}
\vspace{-0.3cm}
\end{table}

\subsubsection{The importance of each feature} To test the importance of each input feature, all features are used as input to the model one by one. The dataset has the same SNR and sample time as Table~\ref{Table: The scalability of intra links}.

Table~\ref{Table: The importance of each feature} shows the test accuracy when a single feature is used as the input of the classifier. 
% Different from the high-resolution data in~\cite{sun2023public}, 
The average test accuracy of low-resolution data is extremely low. However, the stator voltage and current are still a good choice for phase-related fault classification since they could reflect the main fault features. Since the rotor current is a DC signal, the performance of the model based on it is worse than that of the stator voltage and current. Consistent with other studies, although LSTM is designed to learn time series data, the effect of LSTM alone on feature extraction is not as good as CNN, resulting in the lowest test accuracy in the above experiments.

\begin{table}[!t]
\centering\footnotesize
\caption{The importance of each feature.}
\label{Table: The importance of each feature}
\renewcommand{\arraystretch}{1.3}
\scalebox{1}{\begin{tabular}{ l| l l l l }
\toprule
Models & Stator voltage & Stator current & Rotor current & Speed   \\
\hline
CNN-1 & 52.96\% &53.88\% & 50.27\% & 39.45\%   \\
CNN-2  & 51.88\% & 52.98\% & 50.13\% & 38.78\% \\
CNN-3 & 53.79\% & 55.32\% & 52.96\% & 40.36\% \\
Bi-ResNet & 57.53\% &60.86\% & 55.04\% & 45.56\%  \\
ResNet18 & 54.85\% &56.79\%  & 53.85\% &42.06\%  \\
CNN-4  & 46.09\% & 47.85\% & 42.54\% & 35.54\% \\
LSTM & 11.38\% & 25.45\% & 30.54\% & 24.23\% \\
\bottomrule 
\end{tabular}}
\vspace{-0.3cm}
\end{table}

Based on analytical and comparative experiments, the Bi-ResNet has great potential in synchronous motor electrical fault diagnosis, especially when the dataset is low resolution and noisy. Low resolution will cause high-frequency fault signals to become fragmented and incoherent during sampling, making it difficult for the model to accurately extract features and diagnose faults. Meanwhile, noise will mask the fault signal in raw data mislead the fault diagnosis model and reduce the accuracy of fault diagnosis. The Bi-RseNet can fully take advantage of residual learning to find fault features in incoherent high-frequency signals or distinguish fault signals in a large amount of high-frequency noise for fault diagnosis. Another advantage of Bi-ResNet is extensibility. The simple intra-link mode makes it possible for other models to use intra link layers to improve the accuracy of model fault diagnosis.

\subsection{Ablation Experiments} % 多个内连接 参数的重要性

\subsubsection{The sensitivity of hyper-parameter $n$}
In this subsection, we explore the sensitivity of hyper-parameters in Bi-ResNet on test performance, especially the number of intra-links namely $n$. In models with sequential intra-links, the number of internal high-frequency extractors increases as $n$ increases, which means that more high-frequency components will be decomposed from the raw data.

Table~\ref{Table: The hyper-parameter sensitivity in Bi-ResNet} concludes the test accuracy using the different inner connections in four models including CNN-1, CNN-2, CNN-3, and Bi-ResNet. When $n=0$, Bi-ResNet degenerates into a model similar to ResNet18, resulting in a decrease in accuracy. With the increase of $n$, the test accuracy also increases in all models. Nevertheless, the improvement in accuracy due to intra-links comes at the cost of decoupling the internal computing power at the same layer, which reduces the capacity of each inner signal extractor. When $n=3$, the test accuracy of models remains similar to that when $n=2$. Furthermore, the test accuracy starts to decrease when $n=4$, because the potential computing power is dispersed into useless high-frequency component extraction. Therefore, there is a trade-off between the intra-link number $n$ and the complexity of input signals. The potential rationale behind hyperparameter selection is still on work.
% To elaborate on the rationale behind hyperparameter selection, we need more comprehensive experiments. 
\begin{table}[!t]
\centering\footnotesize
\caption{The hyper-parameter sensitivity in Bi-ResNet.}
\label{Table: The hyper-parameter sensitivity in Bi-ResNet}
\renewcommand{\arraystretch}{1.3}
\scalebox{1}{\begin{tabular}{ l| l l l l l}
\toprule
Models & 0 & 1 & 2 & 3 & 4  \\
\hline
CNN-1 & 95.78\% & 96.89\% & 96.99\% & 96.95\% & 96.75\%  \\
CNN-2  & 94.57\% & 96.54\% & 96.94\% & 96.89\% & 96.14\%  \\
CNN-3 & 96.86\% & 97.32\% & 97.86\% & 97.83\% & 96.96\%  \\
Bi-ResNet & 97.59\% & 99.14\% & 99.32\% & 99.26\% & 98.89\%  \\
\bottomrule 
\end{tabular}}
\vspace{-0.3cm}
\end{table}

\subsubsection{The necessity of spatial-temporal convolution}
In this subsection, we explore the role of spatial-temporal convolutions in Bi-ResNet by replacing them with four 1D convolution layers with $1 \times 3$ kernel size. Table~\ref{Table: The influence of spatio-temporal convolution in Bi-ResNet} shows the test accuracy using Bi-ResNet with and without spatial-temporal convolution blocks. Based on the Table, there is a slight reduction in test accuracy when the spatial-temporal convolution blocks are replaced. A possible explanation is that the intra-link blocks in Bi-ResNet help the model extract high-frequency signals, achieving the similar effect of small convolution kernels. Meanwhile, the large receptive field in spatial-temporal convolution blocks provides a wider range of features for the intra-link blocks.  Although stacking small convolution kernels can achieve the same performance as large convolution kernels, a deeper network is required which is hard to design manually.

\begin{table}[!t]
\centering\footnotesize
\caption{The influence of spatial-temporal convolution in Bi-ResNet.}
\label{Table: The influence of spatio-temporal convolution in Bi-ResNet}
\renewcommand{\arraystretch}{1.3}
\scalebox{1}{\begin{tabular}{ l| l l l l l}
\toprule
Models & With & Without \\
\hline
Bi-ResNet & 99.14\% & 99.07\%   \\
\bottomrule 
\end{tabular}}
\vspace{-0.3cm}
\end{table}

\section{Conclusion}\label{sec:Conclusion}

Deep learning models have achieved outstanding performance in synchronous motor electrical fault diagnosis. Whereas, the inherent problem in motor fault diagnosis is how to automatically extract the unnoticeable high-frequency components from raw data for precise classification. In this light, this paper is dedicated to proposing a novel and elegant shallow neural network by introducing the embedded spatial-temporal convolution block and the intra-links in layers within limited computational resources named Bi-ResNet, learning the potential high-frequency components from motor fault signals for fault diagnosis. Five advanced CNN-based networks and two baselines are compared on high and low-resolution datasets with multi-level noise to verify the efficacy of Bi-ResNet. It is demonstrated to be superior to other networks. The intra-links can also be extended to other models to improve the test accuracy. In ablation experiments, the trade-off trade-off between intra-link number and input data complexity is highlighted.

The key limitation of Bi-ResNet lies in the prior knowledge of the dataset. As mentioned in ablation experiments, too many intra-linked neurons may waste the potential computing power in useless high-frequency component extraction, limiting the test accuracy. Moreover, in another aspect, the manually designed hyper-parameters mean that the model is not optimal. The mode of inner connection can also be explored in further research. Nonetheless, we believe that Bi-ResNet has considerable potential in motor fault classification tasks and other industrial applications.


\begin{thebibliography}{10}
\providecommand{\url}[1]{#1}
\csname url@rmstyle\endcsname
\providecommand{\newblock}{\relax}
\providecommand{\bibinfo}[2]{#2}
\providecommand\BIBentrySTDinterwordspacing{\spaceskip=0pt\relax}
\providecommand\BIBentryALTinterwordstretchfactor{4}
\providecommand\BIBentryALTinterwordspacing{\spaceskip=\fontdimen2\font plus
\BIBentryALTinterwordstretchfactor\fontdimen3\font minus \fontdimen4\font\relax}
\providecommand\BIBforeignlanguage[2]{{%
\expandafter\ifx\csname l@#1\endcsname\relax
\typeout{** WARNING: IEEEtran.bst: No hyphenation pattern has been}%
\typeout{** loaded for the language `#1'. Using the pattern for}%
\typeout{** the default language instead.}%
\else
\language=\csname l@#1\endcsname
\fi
#2}}

\bibitem{bhuiyan2020survey}
E.~A. Bhuiyan, M.~M.~A. Akhand, S.~K. Das, M.~F. Ali, Z.~Tasneem, M.~R. Islam, D.~Saha, F.~R. Badal, M.~H. Ahamed, and S.~Moyeen, ``A survey on fault diagnosis and fault tolerant methodologies for permanent magnet synchronous machines,'' \emph{International Journal of Automation and Computing}, vol.~17, pp. 763--787, 2020.

\bibitem{10151958}
G.~Niu, X.~Dong, and Y.~Chen, ``Motor fault diagnostics based on current signatures: A review,'' \emph{IEEE Transactions on Instrumentation and Measurement}, vol.~72, pp. 1--19, 2023.

\bibitem{10330029}
C.~Gao, B.~Gao, X.~Xu, J.~Si, and Y.~Hu, ``Automatic demagnetization fault location of direct-drive permanent magnet synchronous motor using knowledge graph,'' \emph{IEEE Transactions on Instrumentation and Measurement}, vol.~73, pp. 1--12, 2024.

\bibitem{9580592}
M.~Jiménez-Guarneros, C.~Morales-Perez, and J.~d.~J. Rangel-Magdaleno, ``Diagnostic of combined mechanical and electrical faults in asd-powered induction motor using modwt and a lightweight 1-d cnn,'' \emph{IEEE Transactions on Industrial Informatics}, vol.~18, no.~7, pp. 4688--4697, 2022.

\bibitem{10100645}
A.~Mohammad-Alikhani, B.~Nahid-Mobarakeh, and M.-F. Hsieh, ``One-dimensional lstm-regulated deep residual network for data-driven fault detection in electric machines,'' \emph{IEEE Transactions on Industrial Electronics}, vol.~71, no.~3, pp. 3083--3092, 2024.

\bibitem{cai2021data}
B.~Cai, K.~Hao, Z.~Wang, C.~Yang, X.~Kong, Z.~Liu, R.~Ji, and Y.~Liu, ``Data-driven early fault diagnostic methodology of permanent magnet synchronous motor,'' \emph{Expert Systems with Applications}, vol. 177, p. 115000, 2021.

\bibitem{10472098}
J.~Guo, Q.~He, and F.~Gu, ``Dnocnet: A novel end-to-end network for induction motor drive systems fault diagnosis under speed fluctuation condition,'' \emph{IEEE Transactions on Industrial Informatics}, vol.~20, no.~6, pp. 8284--8293, 2024.

\bibitem{quseiri2024fault}
P.~Quseiri~Darbandeh, ``Fault diagnosis in a permanent magnet synchronous motor using deep learning,'' Ph.D. dissertation, Helmut-Schmidt-Universit{\"a}t/Universit{\"a}t der Bundeswehr Hamburg, 2024.

\bibitem{10319110}
J.~Xie, X.~Zhang, D.~Luo, G.~Qin, and F.~Huang, ``Custom phase space reconstruction image-driven fault diagnosis for pmsm under few-labeled samples,'' \emph{IEEE Transactions on Power Electronics}, vol.~39, no.~2, pp. 2731--2740, 2024.

\bibitem{kingma2022autoencodingvariationalbayes}
\BIBentryALTinterwordspacing
D.~P. Kingma and M.~Welling, ``Auto-encoding variational bayes,'' 2022. [Online]. Available: \url{https://arxiv.org/abs/1312.6114}
\BIBentrySTDinterwordspacing

\bibitem{9750877}
F.~Huang, X.~Zhang, G.~Qin, J.~Xie, J.~Peng, S.~Huang, Z.~Long, and Y.~Tang, ``Demagnetization fault diagnosis of permanent magnet synchronous motors using magnetic leakage signals,'' \emph{IEEE Transactions on Industrial Informatics}, vol.~19, no.~4, pp. 6105--6116, 2023.

\bibitem{10038556}
F.~Parvin, J.~Faiz, Y.~Qi, A.~Kalhor, and B.~Akin, ``A comprehensive interturn fault severity diagnosis method for permanent magnet synchronous motors based on transformer neural networks,'' \emph{IEEE Transactions on Industrial Informatics}, vol.~19, no.~11, pp. 10\,923--10\,933, 2023.

\bibitem{he2016deep}
K.~He, X.~Zhang, S.~Ren, and J.~Sun, ``Deep residual learning for image recognition,'' in \emph{Proceedings of the IEEE conference on computer vision and pattern recognition}, 2016, pp. 770--778.

\bibitem{Yu_2018_CVPR}
X.~Yu, Z.~Yu, and S.~Ramalingam, ``Learning strict identity mappings in deep residual networks,'' in \emph{Proceedings of the IEEE Conference on Computer Vision and Pattern Recognition (CVPR)}, June 2018.

\bibitem{hauser2019residualnetworkslearningperturbation}
\BIBentryALTinterwordspacing
M.~Hauser, ``On residual networks learning a perturbation from identity,'' 2019. [Online]. Available: \url{https://arxiv.org/abs/1902.04106}
\BIBentrySTDinterwordspacing

\bibitem{Sun2018StochasticTO}
\BIBentryALTinterwordspacing
Q.~Sun, Y.~Tao, and Q.~Du, ``Stochastic training of residual networks: a differential equation viewpoint,'' \emph{ArXiv}, vol. abs/1812.00174, 2018. [Online]. Available: \url{https://api.semanticscholar.org/CorpusID:54439624}
\BIBentrySTDinterwordspacing

\bibitem{NEURIPS2022_ecc38927}
\BIBentryALTinterwordspacing
M.~Sander, P.~Ablin, and G.~Peyr\'{e}, ``Do residual neural networks discretize neural ordinary differential equations?'' in \emph{Advances in Neural Information Processing Systems}, S.~Koyejo, S.~Mohamed, A.~Agarwal, D.~Belgrave, K.~Cho, and A.~Oh, Eds., vol.~35.\hskip 1em plus 0.5em minus 0.4em\relax Curran Associates, Inc., 2022, pp. 36\,520--36\,532. [Online]. Available: \url{https://proceedings.neurips.cc/paper_files/paper/2022/file/ecc38927fe5148c66bee64ee8fed1e76-Paper-Conference.pdf}
\BIBentrySTDinterwordspacing

\bibitem{NEURIPS2023_98ed250b}
\BIBentryALTinterwordspacing
P.~Marion, ``Generalization bounds for neural ordinary differential equations and deep residual networks,'' in \emph{Advances in Neural Information Processing Systems}, A.~Oh, T.~Naumann, A.~Globerson, K.~Saenko, M.~Hardt, and S.~Levine, Eds., vol.~36.\hskip 1em plus 0.5em minus 0.4em\relax Curran Associates, Inc., 2023, pp. 48\,918--48\,938. [Online]. Available: \url{https://proceedings.neurips.cc/paper_files/paper/2023/file/98ed250b203d1ac6b24bbcf263e3d4a7-Paper-Conference.pdf}
\BIBentrySTDinterwordspacing

\bibitem{cheng2019high}
B.~Cheng, R.~Xiao, J.~Wang, T.~Huang, and L.~Zhang, ``High frequency residual learning for multi-scale image classification,'' \emph{arXiv preprint arXiv:1905.02649}, 2019.

\bibitem{10485711}
Inderjeet and J.~S. Sahambi, ``Residual frequency content awareness approach for image super resolution,'' in \emph{2024 National Conference on Communications (NCC)}, 2024, pp. 1--6.

\bibitem{fang2024hfgn}
F.~Fang, L.~Hu, J.~Liu, Q.~Yi, T.~Zeng, and G.~Zhang, ``Hfgn: High-frequency residual feature guided network for fast mri reconstruction,'' \emph{Pattern Recognition}, vol. 156, p. 110801, 2024.

\bibitem{10203017}
W.~Purbowaskito, C.-y. Lan, and K.~Fuh, ``The potentiality of integrating model-based residuals and machine-learning classifiers: An induction motor fault diagnosis case,'' \emph{IEEE Transactions on Industrial Informatics}, vol.~20, no.~2, pp. 2822--2832, 2024.

\bibitem{kaya2024efficient}
Y.~Kaya, M.~Kuncan, E.~Akcan, and K.~Kaplan, ``An efficient approach based on a novel 1d-lbp for the detection of bearing failures with a hybrid deep learning method,'' \emph{Applied Soft Computing}, vol. 155, p. 111438, 2024.

\bibitem{chen2021amplitude}
G.~Chen, P.~Peng, L.~Ma, J.~Li, L.~Du, and Y.~Tian, ``Amplitude-phase recombination: Rethinking robustness of convolutional neural networks in frequency domain,'' in \emph{Proceedings of the IEEE/CVF International Conference on Computer Vision}, 2021, pp. 458--467.

\bibitem{zhang1988shift}
W.~Zhang, J.~Tanida, K.~Itoh, and Y.~Ichioka, ``Shift-invariant pattern recognition neural network and its optical architecture,'' in \emph{Proceedings of annual conference of the Japan Society of Applied Physics}, vol. 564.\hskip 1em plus 0.5em minus 0.4em\relax Montreal, CA, 1988.

\bibitem{geron2022hands}
A.~G{\'e}ron, \emph{Hands-on machine learning with Scikit-Learn, Keras, and TensorFlow}.\hskip 1em plus 0.5em minus 0.4em\relax " O'Reilly Media, Inc.", 2022.

\bibitem{bjorck2018understanding}
N.~Bjorck, C.~P. Gomes, B.~Selman, and K.~Q. Weinberger, ``Understanding batch normalization,'' \emph{Advances in neural information processing systems}, vol.~31, 2018.

\bibitem{8842598}
R.~Liu, F.~Wang, B.~Yang, and S.~J. Qin, ``Multiscale kernel based residual convolutional neural network for motor fault diagnosis under nonstationary conditions,'' \emph{IEEE Transactions on Industrial Informatics}, vol.~16, no.~6, pp. 3797--3806, 2020.

\bibitem{li2023fault}
Z.~Li, P.~Wang, and X.~Li, ``Fault diagnosis of asynchronous motors based on 2dcnn-gru network optimization,'' in \emph{2023 CAA Symposium on Fault Detection, Supervision and Safety for Technical Processes (SAFEPROCESS)}.\hskip 1em plus 0.5em minus 0.4em\relax IEEE, 2023, pp. 1--6.

\bibitem{sun2023public}
Z.~Sun, R.~Machlev, Q.~Wang, J.~Belikov, Y.~Levron, and D.~Baimel, ``A public data-set for synchronous motor electrical faults diagnosis with cnn and lstm reference classifiers,'' \emph{Energy and AI}, vol.~14, p. 100274, 2023.

\end{thebibliography}
\end{document}